\documentclass[aps,prl,twocolumn,showpacs]{revtex4}

\usepackage{amssymb}
\usepackage{amsmath}
\usepackage[pdftex]{graphicx}
\providecommand{\unit}[1]{\,\mbox{#1}} % UNIDADES

\begin{document}

\title{Many-body effects on the $\rho_{xx}$ ringlike structures in two-subband wells}
\author{Gerson J.~Ferreira}
\author{Henrique J.~P.~Freire}
\author{J.~Carlos Egues}
\affiliation{Departamento de F\'{\i}sica e Inform\'{a}tica, Instituto de F\'{\i}sica de S\~{a}o Carlos, Universidade de S\~{a}o Paulo, 13560-970 S\~{a}o Carlos, S\~{a}o Paulo, Brazil}
\date{\today}

\begin{abstract}
The longitudinal resistivity $\rho_{xx}$ of two-dimensional electron gases formed in wells with two subbands
displays ringlike structures when plotted in a density--magnetic-field diagram, due to
the crossings of spin-split Landau levels (LLs) from distinct subbands. Using spin
density functional theory and linear response, we investigate the shape and spin polarization of these structures as a function of 
temperature and magnetic-field tilt angle. We find that (i) some of the rings ``break''
at sufficiently low temperatures due to a quantum Hall ferromagnetic phase transition, thus
exhibiting a high degree of spin polarization ($\sim 50 $\%) within, consistent with the NMR
data of Zhang \textit{et al.} [Phys. Rev. Lett. {\bf 98}, 246802 (2007)], and (ii) for
increasing tilting angles the interplay between the anticrossings due to inter-LL couplings
and the exchange-correlation (XC) effects leads to a collapse of the rings at some critical angle $\theta_c$, in agreement with the data of
Guo \textit{et al.} [Phys. Rev. B {\bf 78}, 233305 (2008)].
\end{abstract}

\pacs{73.43.Qt,71.15.Mb,73.43.Nq}
\maketitle

The fascinating quantum Hall regime hosts a number of fundamental
physical phenomena, being also relevant for metrology
(standard for resistance) and as an alternate means to precisely determine the fine structure
constant. The spectrum of two-dimensional electron gases (2DEGs) in the quantum Hall regime
is quantized into highly degenerate Landau levels (LLs) \cite{Chakraboty95}. At opposite-spin LL crossings near
the Fermi level, a ferromagnetic instability of the 2DEG may arise thus leading to a quantum Hall
ferromagnetic phase. This spontaneous spin polarization of the electrons lowers the repulsive Coulomb energy
of the Fermi sea, because electrons with parallel spins avoid each other due the Pauli exclusion principle. Even for
vanishingly small Zeeman splittings the exchange-energy gain can stabilize quantum Hall ferromagnetism at low enough
temperatures \cite{Quinn85,GirvinBook1997,Jungwirth00}.

\begin{figure}[b!]
   \centering
   \includegraphics[width=7.8cm,keepaspectratio=true]{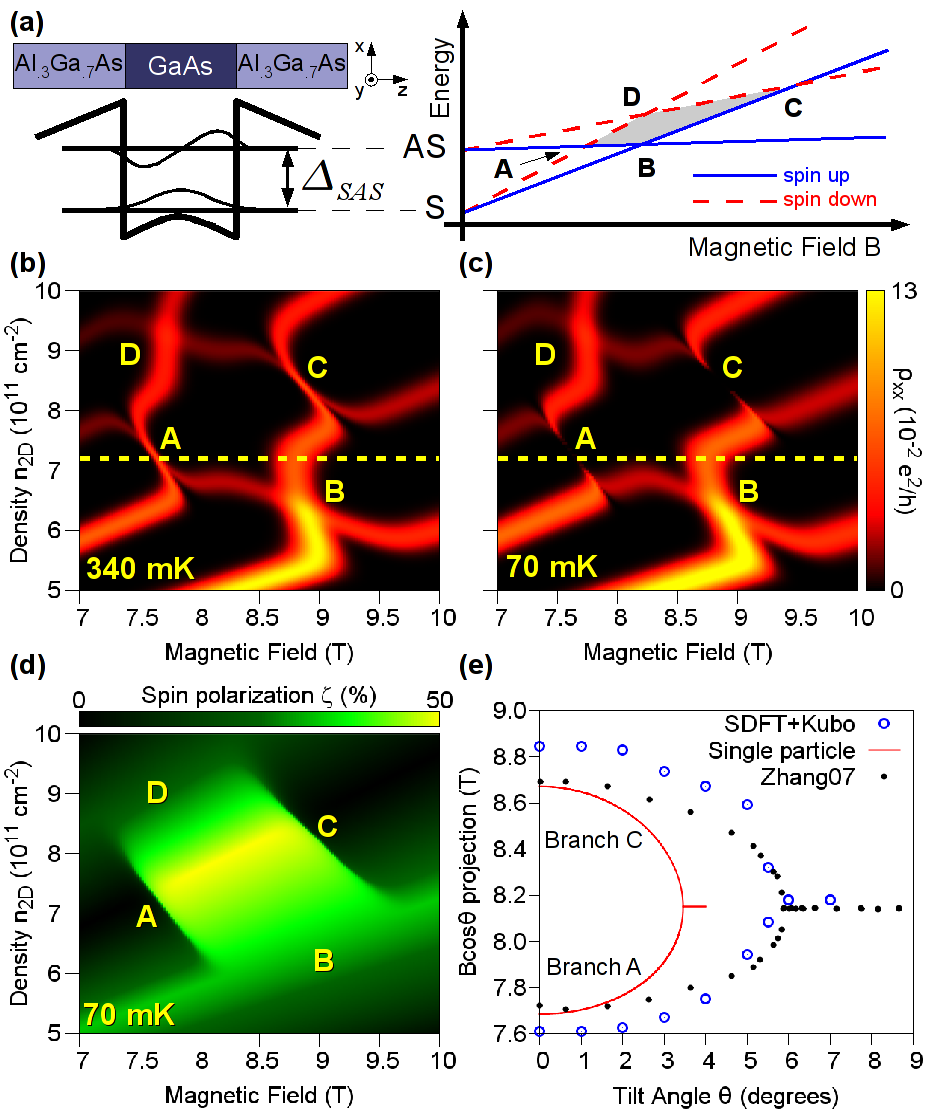}
   \caption{(a) Two-subband GaAs well and schematic fan diagram with
LL crossings from distinct subbands. The ABCD loop gives rise to
ringlike structures in the calculated $n_{\text{2D}}$--$B$
diagram of $\rho_{xx}$ (b) for $\nu=4$. This ring ``breaks''
at lower temperatures (c) due to quantum Hall ferromagnetic transitions, thus displaying a
high spin polarization within (d) \cite{Zhang06_PRL}.
For increasing $B$ field tilt angles $\theta$, the ring shrinks (i.e.,
A and C move closer) and fully collapses (A=C) at $\theta_c$ (e),
in agreement with the data \cite{ZhangSU4-07} (cf. empty and solid circles).}
   \label{fig1}
\end{figure}

Quantum Hall ferromagnetism has been extensively studied in the quantum
Hall regime via magnetotransport measurements \cite{qhf-exp}.
Near LL crossings in tilted magnetic fields $B$, the longitudinal resistivity
$\rho_{xx}$ vs $B$ of wells with a singly occupied subband exhibits ubiquitous
hysteretic spikes, which signals quantum Hall
ferromagnetism \cite{spikes,jungwirthprl,Freire07PRL}. In two-subband wells, spin-split LLs from
distinct subbands cross even without a tilted $B$ field and can form closed loops [ABCD loop, Fig.~\ref{fig1}(a)].
Quite generally, $\rho_{xx}$ is directly related to the energy spectrum near the Fermi level (linear response) and hence
the topology in Fig. 1(a) translates into ringlike structures \cite{Muraki01,Zhang05} in $\rho_{xx}$ when plotted in a
density--$B$-field diagram $n_{\text{2D}}-B$, Fig.~\ref{fig1}(b). Recently Zhang \textit{et al.} \cite{Zhang06_PRB} have shown that
near opposite-spin LL crossings the rings ``break'' at low enough
temperatures ($70 \unit{mK}$). NMR measurements \cite{Zhang06_PRL} near the broken edge C show a
high degree of spin polarization, which points to a ferromagnetic instability of the 2DEG.
For tilted $B$ fields some of rings ``shrink'', fully
collapsing for angles above a critical value \cite{ZhangSU4-07}.

Here we use spin density functional theory (SDFT) \cite{dft} together with a linear response
model \cite{ando-uemu-gerh} to investigate the shape and spin polarization
of the ringlike structures in realistic quantum wells with two subbands
at various filling factors $\nu$, as a function of the temperature and tilt angle $\theta$ of the
$B$ field. We find that exchange-correlation (XC) effects are crucial to \textit{quantitatively}
describe the experiments: in particular (i) the $\nu=4$ ring breakup at low temperatures [Fig.~\ref{fig1}(c)] follows from quantum Hall
ferromagnetic phase transitions \cite{Zhang05,Zhang06_PRB}, Fig. 2.
The calculated spin polarization [Fig.~\ref{fig1}(d)] within the
broken ring reaches $50\%$, consistent with NMR data \cite{Zhang06_PRL}. (ii) The shrinkage of the $\nu=4$
ring for increasing $\theta$ and its full collapse at $\theta = \theta_c$  [Fig.~\ref{fig1}(e)]  \cite{ZhangSU4-07} arise from the
interplay between the anticrossings due to the inter-LL couplings and the exchange field, Fig. 3.
We note that only rings formed from consecutive LLs, for which inter-LL coupling is
operative, collapse for increasing $\theta$.

The quantum phase transitions
we find here are not specific to the $\nu=4$ ring. They are general and
should also occur for $\nu=6$ and others, but for distinct ranges of parameters. Other
2DEG systems, e.g., formed in Mn-based wells \cite{mn-wells}, can also show peculiar
ring structures.

\textit{System.} We consider the structures of Zhang \textit{et al.} \cite{Zhang06_PRL,Zhang05,Zhang06_PRB,ZhangSU4-07}: a
wide $240\unit{\AA}$ GaAs square quantum well with Al$_{0.3}$Ga$_{0.7}$As
barriers and symmetric $\delta$-doping (Si) with $240\unit{\AA}$ spacers [Fig.~\ref{fig1}(a)].
The electron density in the well is controlled by a gate voltage, as in an ideal capacitor model \cite{Yamaguchi06}.
At zero bias $n_{\text{2D}} = 8.1 \times 10^{11} \unit{cm}^{-2}$. The electron mobility $\mu_e = 4.1 \times 10^5
\unit{cm}^2/\unit{Vs}$ is assumed constant for the entire $B$ field and gate voltage ranges.

\textit{Kohn-Sham problem.} The Kohn-Sham implementation
of density-functional theory maps the problem of fully
interacting electrons onto a non-interacting Schroedinger equation
-- the KS equation -- with electrons in an effective single-particle potential \cite{dft}.
For magnetic fields $B$ tilted $\theta$ with the 2DEG normal, this reads
\begin{equation}
  (H_{\parallel} + H_{z}^{\sigma_z} + \delta H_{\theta})\psi = \varepsilon\psi,
\label{eq1}
\end{equation}
with
\begin{eqnarray}
 %\begin{array}{r c l}
   H_{\parallel} &=& \dfrac{P_x^2}{2m}+\dfrac{1}{2}m\omega_c^2(x-x_0)^2, \label{eq2} \\
   H_{z}^{\sigma_z} &=& \dfrac{P_z^2}{2m}+\dfrac{1}{2}m\omega_p^2 z^2+ \dfrac{1}{2}g_e\mu_B \sigma_z B+v_{eff}^{\sigma_z}(z), \label{eq3} \\
   \delta H_{\theta} &=& \omega_p z P_x, \label{eq4}
\end{eqnarray}
where $m$ ($0.067m_0$) is the effective mass, $g_e$ ($-0.44$) the bulk g-factor, $P_{x,y,z}$ the $x,y,z$ components of the
electron momentum operator, $\omega_c = eB\cos\theta/m$
the cyclotron frequency, $\sigma_z=\pm$ (or $\uparrow,\downarrow$), $\omega_p = eB\sin\theta/m$, $x_0 = - \ell_0^2 P_y/\hbar$,
$\ell_0^2 = \hbar/eB\cos\theta$ the magnetic length and
%$v_{eff}^{\sigma_z}(z)$ is given by
\begin{equation}
v_{eff}^{\sigma_z}(z) = v_c(z) + v_H(z; [n]) +
v_{xc}^{\sigma_z}(z; [n_{\uparrow},n_{\downarrow}]), \label{eq5}
\end{equation}
$v_c(z)$ is the structural well potential. The
Hartree potential $v_H(z; [n])$ is obtained self-consistently from Poisson's equation.
For the XC potential $v_{xc}^{\sigma_z}(z; [n_{\uparrow},n_{\downarrow}])$, we use the PW92 parametrization \cite{PW92} of the local-spin-density approximation (LSDA) \cite{AXC}.
Here we have approximated the electron density $n(x,y,z)$ by its average over the $xy$ plane $n(z)=n_\uparrow(z)+n_\downarrow(z)$ \cite{Freire07PRL}. This renders both the Hartree and the XC potentials dependent upon only $z$.

\textit{Perpendicular $B$ field}. For $\theta=0^\circ$, $\omega_p = 0 \Rightarrow\delta H_{\theta}=0$, the KS equation \eqref{eq1} is separable in the $xy$ and $z$
variables and has eigenfunctions  $\psi_{i,n,k_{y}}^{\sigma_{z}}(x,y,z) = \frac{1}{\sqrt{L_y}} \exp(\mathrm{i}k_{y}y) \varphi_{n}(x) \chi_{i}^{\sigma_{z}}(z)$ (Landau gauge), with $\varphi_{n}(x)$ being the \textit{n}-th harmonic-oscillator
eigenfunction centered at $x_{0}=-\hbar k_{y}/m\omega_{c}$ and $k_{y}$
the electron wave number along the $y$ axis; $L_{y}$ is a
normalizing length. The KS eigenenergies are
$\varepsilon_{i,n}^{\sigma_z}=\varepsilon_{n} +\varepsilon_i^{\sigma_z}+g_e\mu_B \sigma_z B/2$ (``Landau fan diagram''), with
$\varepsilon_{n} = \left(n+1/2\right)\hbar\omega_c$, $n = 0, 1,...$ the LL energies (degeneracy $n_B=eB/h$) and $\varepsilon_i^{\sigma_z}$ the quantized levels obeying $H_z^{\sigma_z}\chi_i^{\sigma_z}=\varepsilon_i^{\sigma_z}\chi_i^{\sigma_z}$, $i=0,1,...$. with a self-consistently calculated
chemical potential $\mu$.

\textit{Tilted $B$ field.} For $\theta > 0^\circ$ the KS equation \eqref{eq1}
is not separable because $\delta H_{\theta} \sim \sin\theta zP_x \neq 0$. However,
since $\delta H_{\theta} \ll H_0 = H_{\parallel}+H_{z}^{\sigma_z}$ and only couples consecutive
LLs from distinct subbands, we can obtain the KS solutions $\tilde {\psi}(x,y,z;\theta)$
as an expansion in terms of the eigenfunctions  $\phi_{i,n,k_{y}}^{\sigma_{z}}(x,y,z;\theta)$ of $H_0$.
% \footnote{Note that this is different from the $\theta = 0$ case, since here $\omega_p \neq 0$ in $H_{z}^{\sigma_z}$.}
We perform this expansion \textit{at every iteration} of our self-consistent scheme.
We obtain good results by truncating the expansion for energies greater than
$\mu + k_BT$. This LL coupling leads to
anticrossings of the KS energies for equal-spin LLs,
which ultimately make the ring shrink for tilted fields, Fig. 3.

\textit{Linear-response $\rho_{xx}$}. By \textit{assuming} that the KS eigenvalues
$\varepsilon_{i,n}^{\sigma_z}$ represent the eigenenergies of the actual (Fermi-liquid) quasi-particles in our 2DEG, we
use them in a Kubo-type formula \cite{ando-uemu-gerh}
to calculate the conductivity tensor $\sigma$. For instance, within the self-consistent
Born-approximation with short-range scatterers \cite{ando-uemu-gerh}
$\sigma_{xx} = \frac{e^2}{\hbar \pi^2} \int\limits_{-\infty}^{\infty}
\left( -\frac{\partial f(\varepsilon)}{\partial \varepsilon} \right)
\notag \sum_{i,n,\sigma_z} \left(n+\frac{1}{2}\right)\exp \left[ -
\left( \frac{\varepsilon-\varepsilon_{i,n}^{\sigma_z}}
{\Gamma_{\mathrm{ext}}} \right)^2 \right] \mathrm{d} \varepsilon$,
$\Gamma_{ext}$ denotes the width of the extended-state region
% \footnote{The width of the extended states region $\Gamma_{ext}$ affects only qualitatively our results.}
within the broadened density of states
and $f(\varepsilon)$ the Fermi function. We obtain the resistivity from $\rho=\sigma^{-1}$.

\textit{Spin-polarized rings.} Figures \ref{fig1}(b) and \ref{fig1}(c) show our calculated $n_{\text{2D}}-B$ diagram of $\rho_{xx}$ for two different temperatures $T = 340\unit{mK}$ and $T = 70\unit{mK}$, respectively, near the $\nu=4$ ring. Similarly to the experiment of Ref.~\cite{Zhang06_PRB}, we find that the $\nu=4$ ring ``breaks'' at the opposite spin LL crossings (points \textit{A} and \textit{C}) at lower temperatures \cite{intensity}, Fig.~\ref{fig1}(c). Figure \ref{fig1}(d) shows the corresponding $n_{\text{2D}}-B$ diagram of the spin-polarization $\xi=(n_{\text{2D}}^\uparrow - n_{\text{2D}}^\downarrow)/n_{\text{2D}}$, $n_{\text{2D}}=n_{\text{2D}}^\uparrow + n_{\text{2D}}^\downarrow$. Interestingly, we find a high spin polarization ($\sim 50$ \%) within the $\nu=4$ ring. This high polarization and, more importantly, its abrupt variation at the opposite-spin crossings \textit{A} and \textit{C} signal  quantum Hall ferromagnetic phase transitions. For $T = 340\unit{mK}$ the spin polarization of the $\nu=4$ ring (not shown), though high, varies smoothly at the opposite spin crossings. We note that the high spin polarization $\xi$ within the ring points to quantum Hall ferromagnetism, being also consistent with resistively-detected NMR data available \cite{Zhang06_PRL}; however, we contend that the high $\xi$ \textit{and} the discontinuities of $\rho_{xx}$ at the crossings A and C [Fig. 1(c)] constitute the signature for the quantum Hall ferromagnetic instability.
\begin{figure}[t!]
   \centering
   \includegraphics[width=8cm,keepaspectratio=true]{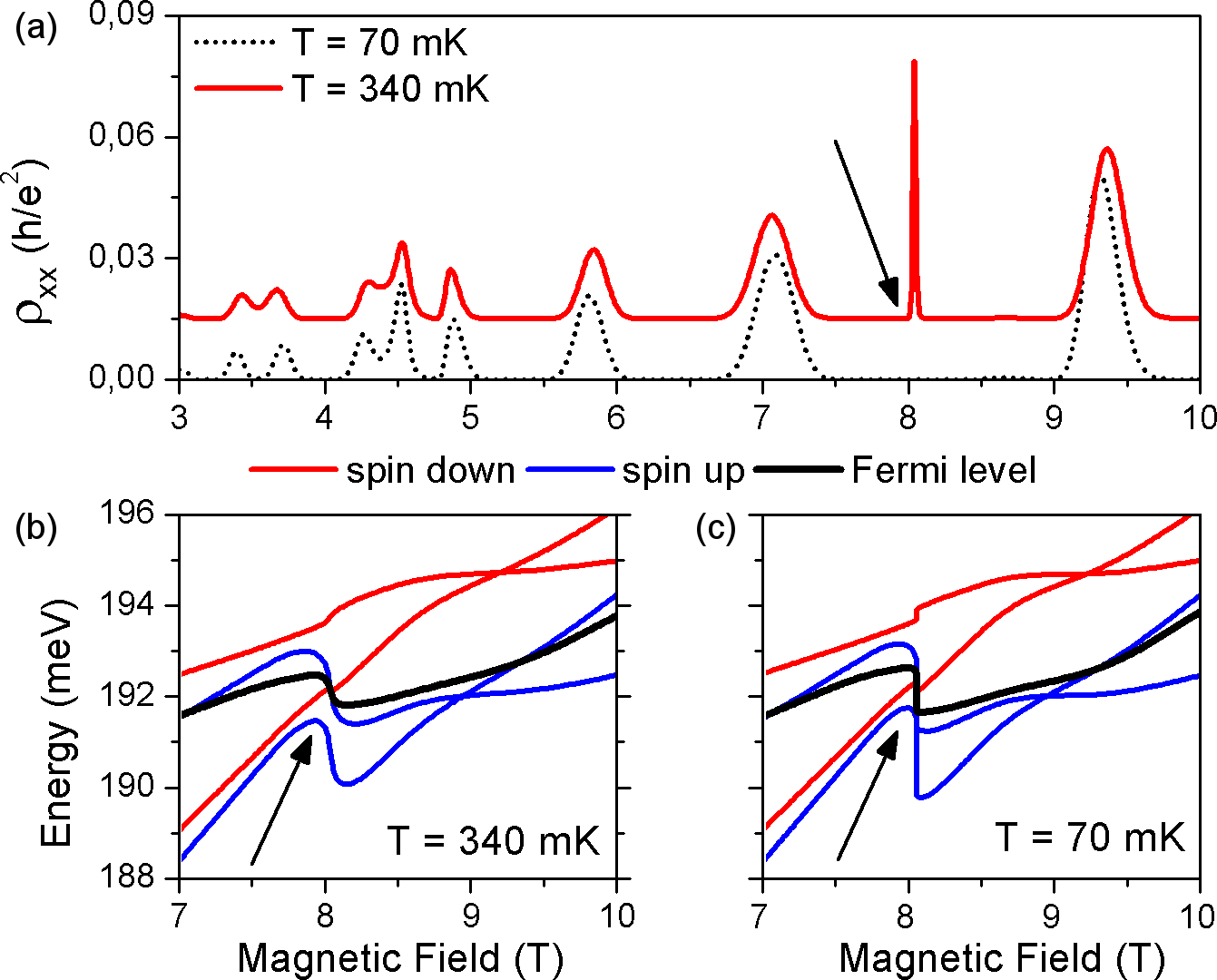}
   \caption{(a) Longitudinal resistivity $\rho_{xx}$ calculated for $n_{\text{2D}} = 7.3 \times 10^{-11}\unit{cm}^{-2}$ [see horizontal dashed line in Figs. 1(b)-1(c)] for $T = 70 \unit{mK}$ (dotted line) and $T = 340 \unit{mK}$ (solid line, slightly shifted upwards for clarity). Note the distinctive spike in the higher temperature  $\rho_{xx}$ at $B \approx 7.6 \unit{T}$. (b) and (c) show the corresponding Landau fan diagrams for low and high temperatures, respectively. Note that for $T = 70 \unit{mK}$, the LLs show discontinuities near $B = 7.6 \unit{T}$. These discontinuities suppress the $\rho_{xx}$ spike at $B \approx 7.6\unit{T}$ because the corresponding levels do not actually cross the chemical potential $\mu$ -- they suddenly jump over it. At higher  temperatures the Landau diagram is continuous and the spike appears in $\rho_{xx}$.}
   \label{fig2}
\end{figure}

The contrast between the low and high temperature results is more clearly seen in Fig.~\ref{fig2}, which shows $\rho_{xx}$ for $n_{\text{2D}} = 7.3 \times 10^{11}\unit{cm}^{-2}$ at $340\unit{mK}$ and $70\unit{mK}$. The spike near $B = 7.6\unit{T}$ [see arrow in Fig.~\ref{fig2}(a)] comes from the left edge of the $\nu = 4$ ring (point \textit{A} in Fig.~\ref{fig1}) and is suppressed at $T = 70\unit{mK}$. The Landau fan diagrams for both temperatures differ substantially only around this region [see arrows in  Figs.~\ref{fig2}(b)-(c)]. At $T = 70\unit{mK}$ the diagram shows an abrupt transition and the chemical potential $\mu$ jumps to the spin-down state of the lower subband, thus suppressing the $\rho_{xx}$ spike. Note also the exchange enhancement of the spin splittings in Fig. 2(b)-(c) when $\mu$ lies essentially between the spin-split LLs.

A relevant parameter in our simulations is the LL broadening $\Gamma$. For short-range scatterers, the electron mobility $\mu_e$ and the LL broadening are related by $\Gamma= \Gamma_0 \sqrt{B/\mu_e}$ \cite{Ando83}, with $\Gamma_0 = (2/\pi)^{1/2}\hbar e/m$. We use $\Gamma_{70} = 0.130\sqrt{B}\unit{meV}$ and $\Gamma_{340} = 0.150\sqrt{B}\unit{meV}$ to simulate the ring structures at $T = 70\unit{mK}$ and $340 \unit{mK}$, respectively [see Figs.~\ref{fig1}(b)-(c)] \cite{broadening}. Note that the temperature-dependent $\Gamma_0$  differs from the one determined from the zero voltage $\mu_e$, $\Gamma = 0.210\sqrt{B} \unit{meV}$.
A strong dependence of $\mu_e$ on the gate voltage (or density) is reported in \cite{EllenbergerPHD} for parabolic two-subband wells, which also show ringlike structures \cite{Ellenberger06,EllenbergerPHD}.
Hence treating $\Gamma_0$ as a fitting parameter is somewhat justifiable here.

\begin{figure}
   \centering
   \includegraphics[width=8 cm,keepaspectratio=true]{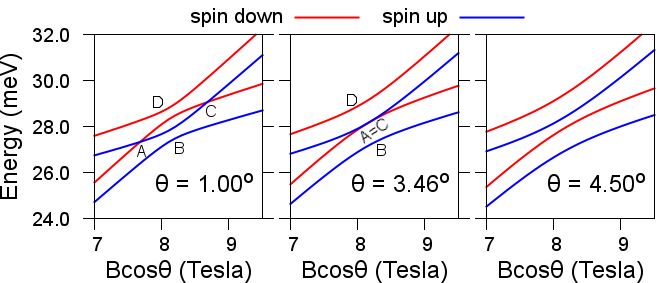}
   \caption{Non-interacting (a)-(b) and interacting (c)-(d) Landau fan diagrams near the $\nu = 4$ ring for several tilt angles, showing the anti-crossings (B \& D) due to the inter-LL coupling, Eq. (\ref{eq4}). For increasing $\theta$ the anti-crossings near D \& B become larger thus making A \& C move closer and the ring shrink. The non-interacting model \cite{non-int-model} gives a qualitative picture of the ring collapse, while our many-body calculation (c)-(d) ($T = 70 \unit{mK}$) provides a quantitative description and shows the relevance of competing XC effects, see Fig. 1(e).}
   \label{fig3}
\end{figure}

\textit{Ring shrinking and collapse.} In the experiment of Ref.~\cite{ZhangSU4-07} the authors show the projection of the $\rho_{xx}$ side crossings [points A \& C in Fig.~\ref{fig1}] onto the $B_{\bot} = B\cos{\theta}$ axis as a function of $\theta$. In Fig.~\ref{fig1}(e) we compare the  experimental data (black dots) with the results of a non-interacting model (solid line) and our SDFT + Kubo approach (circles). The solid line is obtained from a non-interacting model (discussed in detail in Ref. \cite{non-int-model}) with effective parameters that fit the $\theta = 0^\circ$ data. This simple  model  illustrates qualitatively the effects of the inter-LL coupling on the ring collapse, Fig.~\ref{fig3}(a)-(b):  the anticrossings [near D \& B] between LLs with the same spin increase with $\theta$, thus making the opposite spin LL crossings [points A \& C] move closer in energy, effectively shrinking the ring. Note, however, that the collapsing angle for this non-interacting model ($3.46^{\circ}$) is about half of the experimental value  $\theta_c^{exp} \approx 6^{\circ}$ \cite{ZhangSU4-07}.

Our many-body calculation, on the other hand, agrees well with the data
[empty circles, Fig.~\ref{fig1}(e)] \cite{footnote-error}. Here,
as the tilt angle increases, a competition sets in between the inter-LL coupling and the exchange-enhanced spin splittings of the LLs. While the inter-LL coupling tends to shrink the rings [anticrossings, Figs. 3(a)-(b)], the enhanced spin splittings make the rings larger [e.g., Figs. 2(b)-(c) with no inter-LL coupling (i.e., $\theta=0^\circ$)].
This interplay ``delays'' the ring collapse, Figs. 3(c)-(d); here we find $\theta_c^{theory} \approx 6^\circ$ -- in agreement with \cite{ZhangSU4-07}.

\textit{Some remarks.} Experimentally \cite{Zhang06_PRB}
all crossings of the $\nu = 4$ ring are broken at low temperature [points A, B, C \& D].
In our approach the quantum Hall phase transitions occur only at opposite-spin LL crossings
[points A \& C]. Interestingly, we note that the inter-LL coupling makes the ring break near the same-spin LL crossings
even for very small angles [see D \& B in Figs. 3(c)-(d)]. In Ref. \cite{non-int-model} we have also obtained the $n_{\text{2D}}-B$
map of $\rho_{xx}$ for a non-interacting model and find ring breakups near D \& B.
However, the actual ring breakups near D \& B at $\theta=0^\circ$ could also be related to the derivative discontinuity of the XC
functionals \cite{Leeor2008}, which is absent in local (LSDA) and semi-local (e.g., generalized gradient approximations)
functionals \cite{Perdew1983, Sham1983}. This discontinuity
appears as a jump in $n_{\text{2D}}$ at the threshold for the second subband occupation at $B=0$ \cite{ProettoPRL07}. The results of Ref.~\cite{ProettoPRL07} suggest that orbital functionals (e.g., exact-exchange) \cite{Leeor2008} may give rise to phase transitions
at the crossings B \& D. Clearly more work is needed here. A Hartree-Fock analysis in
model bilayer systems \cite{Jungwirth00} suggests that quantum Hall ferromagnetic
instabilities can occur at same-spin pseudospin LL crossings \cite{Jungwirth00}.

\textit{Summary.} We have combined SDFT and linear response to investigate magnetotransport in 2DEGs formed in
two-subband wells \cite{Zhang06_PRB,Zhang06_PRL,ZhangSU4-07}. Our calculated $n_{\text{2D}}-B$ maps
of $\rho_{xx}$ show ringlike structures. At low temperatures the $\nu=4$ ring breaks due to
quantum Hall ferromagnetic phase transitions. The $n_{\text{2D}}-B$ diagram of the 2DEG spin polarization
$\xi$ shows the ring to be $\sim 50$\% spin polarized. For tilted $B$ fields, the $\nu=4$ ring
shrinks and fully collapses at a critical angle $\theta_c\approx 6^\circ$, in excellent agreement with the
data \cite{ZhangSU4-07}. The interplay between the equal-spin LL anti-crossings and the XC effects are crucial here.
%The good agreement with recent data \cite{Zhang06_PRB,Zhang06_PRL,ZhangSU4-07} suggests that our SDFT + Kubo approach can
%be used to study some aspects of magnetotransport in the quantum Hall regime.
A direct experimental evidence of our prediction of a high $\xi$ in the ring is still lacking; the resistively-detected NMR data of Ref. \cite{Zhang06_PRL} only shows signals near point C.  We hope our work stimulates further investigations in the literature.

\textit{Acknowledgments.}
GJF acknowledges useful conversations with X.~C.~Zhang and T.~Ihn.
This work was supported by FAPESP and CNPq.


\begin{thebibliography}{36}
\bibitem{intensity} The $(n+1/2)$ ``short range'' factor  \cite{ando-uemu-gerh} in $\sigma_{xx}$ enhances $\rho_{xx}$ on 
the $n_{\text{2D}}-B$ plane for higher LL indices $n$. We have also simulated scatterers of slowly varying type by removing this factor. Our
results are essentially unaltered.
\bibitem{broadening} Following Ref. \cite{Ando83}, we 
find that near LL crossings the density of states deviates somewhat from the
gaussian of individual LLs; however, our results do not change 
in any significant way in the range of parameters studied.
\bibitem{footnote-error} The error in the SDFT + Kubo calculation [Fig. 1(e)] arises
not only from the approach used, but also from extracting ``by inspection''
the distance AC from Fig. 1(c).
\end{thebibliography}
\end{document}